# Premature Switching in Graphene Josephson Transistors

Feng Miao[‡], Wenzhong Bao[‡], Hang Zhang and Chun Ning Lau*

Department of Physics and Astronomy, University of California, Riverside, Riverside, CA 92521

## Abstract

We investigate electronic transport in single layer graphene coupled to superconducting electrodes. In these Josephson transistors, we observe significant suppression in the critical current $I_c$ and large variation in the product $I_c R_n$ in comparison to theoretic predictions in the ballistic limit. We show that the depression of $I_c$ can be explained by premature switching in underdamped Josephson junctions described within the resistively and capacitively shunted junction (RCSJ) model. By considering the effect of premature switching and dissipation, the calculated gate dependence of product $I_c R_n$ agrees with experimental data. Our discovery underscores the crucial role of thermal fluctuations in electronic transport in graphene Josephson transistors.

[‡]These authors contributed equally.
*To whom correspondence should be addressed. Email: lau@physics.ucr.edu

Since its first discovery in 2004 [1], graphene- a monolayer of carbon atoms arranged in a honey-comb lattice- has fascinated both theorists and experimentalists in condensed matter physics. Because of its unique linear dispersion relation [2-4], many novel phenomena have been predicted to arise in graphene, such as Veselago lensing effect [5], spin Hall Effect [6] and half-metallic ribbons [7]. Moreover, as a surface two dimensional (2D) electron system, graphene can be easily coupled to special electrodes such as superconductors with controllable interfaces, while its charge density and type are continuously tunable by an electrostatic gate. Thus, coupling graphene to superconducting electrodes provides an ideal platform for investigating the interplay between superconductivity and relativistic quantum electrodynamics, which will be useful for understanding of high temperature and heavy-element superconductors[8]. Superconducting order is also predicted to emerge in pure and doped graphene[9].

Coupling superconducting electrodes to graphene yields a Josephson junction (JJ) [10]. In contrast to traditional JJ, in which the weak links are insulators, normal metals (N) or constrictions, a superconductor/graphene/superconductor (SGS) JJ consists of a ballistic, 2D relativistic system with gate-tunable conductance [11-14]. Several unusual phenomena were predicted to arise in this novel system, such as specular Andreev reflection [15], novel propagating modes of Andreev electrons [16] and oscillation of tunneling probability with barrier width [17]. Experimentally, the observation of bi-polar supercurrent and multiple Andreev reflection (MAR) in graphene Josephson junctions has been reported by several groups [18-20]. Despite these initial studies, much remains to be explored. In particular, the observed critical current $I_c$ is much reduced from the theoretical value, and the $I_c R_n$ product (where $R_n$ is the normal state resistance of the junction) display a much larger dependence on graphene's charge density than what one would expect for ballistic SGS junctions [21]. This discrepancy was tentatively attributed to disorder[22], but no consensus has been reached.

In this paper, we report the observation of reduction of critical current $I_c$ and large variation of gate tunable product $I_c R_n$ in SGS junctions. The results are considered within the framework of resistively and capacitively shunted junctions (RCSJ) [23]. We show that both the reduced $I_c$ and large gate dependence of $I_c R_n$ product can be accounted for by dissipation and premature switching in underdamped junctions [21] at a finite temperature. Our results indicate the important role of thermal effect in electronic transport in SGS junctions.

The SGS junctions are fabricated from mechanical exfoliation of single layer graphene onto $SiO_2$/Si wafers, which is degenerately doped to act as back gates. The single layer graphene is identified by color contrast in an optical microscope. After electron beam lithography that defines the electrode patterns, we deposit 5nm titanium and 80nm aluminum. Here the Ti layer acts as an adhesion layer, and Al is a superconductor. Fig. 1a shows an SEM image of a typical device. The source-drain separations of the SGS devices vary between 200nm and 550nm. All measurements are taken in $He^3$ low temperature refrigerator with base temperature of 300mK. A copper-powder filter and two RC filters [24] are integrated in the circuit to ensure low

Fig. 1b displays the differential conductance for a typical device as a function of voltage across the junction for $V_g$=0V. The giant center peak at $V$=0 indicates the proximity-induced supercurrent. The peaks at $V=\pm 202$ µV mark the onset of quasiparticle conductance and correspond to $2\Delta$, where $\Delta$ is the superconductor energy gap. Thus, for this SGS junction, we infer $\Delta$ =101 µeV, which is smaller than $\Delta$=180 µeV for bulk Al. For our other devices, $\Delta$ ranges from 90 to 120 µV. At $V<2\Delta$, there are a series of smaller conductance peaks; such sub-gap features, which are commonly observed in superconductor/normal metal/superconductor (SNS) junctions, arise from the phenomenon of multiple Andreev reflection in conductance. During an Andreev reflection, an electron in the normal metal incident on the N/S interface is reflected as a hole, while transferring $2e$ to the superconducting Cooper pair condenstate [25]. In SNS junctions, an electron in graphene can be reflected back and forth several times between the two N/S interfaces, gaining energy $eV$ each time it transverses the junction, until it accumulates sufficient energy to exit graphene into superconducting electrode as a quasiparticle. Thus multiple Andreev reflection (MAR) gives rise to features [26] at biases which are sub-harmonic values of $2\Delta$, *i.e.,* at $V=2\Delta/ne$, where $n$ is an integer and $e$ is the electron charge. From Fig. 1b, the sub-gap features occur at V=100, 68.7 and 47 µV, corresponding to $n$=2, 3 and 4, respectively.

We now focus on the voltage-current (*V-I*) characteristics of the graphene Josephson junctions. We current bias the device and monitor the voltage $V$ across the SGS junction dropped on graphene. Fig. 2a and 2c plots $V$ (color) vs. current bias $I_b$ (vertical axis) and gate voltage $V_g$ (horizontal axis) for two different devices. The separation $L$ between two electrodes of this device is about 260nm and the aspect ratio $W/L$ of graphene is ~ 10 ($W$ is the width of the conduction channel). Supercurrent is observed in all devices (>10) with transparent graphene/superconducting electrode interfaces. The observed *V-I* curves are strongly dependent on the applied gate voltages (line traces at different $V_g$ are shown in Fig 2b and 2d). Strikingly, sueprcurrent is observed in electron-doped as well as hole-doped regimes, and even at the charge neutrality point which has nominally zero charge density. The values of critical current $I_c$, at which the junction switches from zero resistance state to a resistive state, are also strongly gate dependent. For the device shown in Fig. 2a, $I_c$ ~ 200nA at the Dirac point and ~450nA when highly doped. Similar gate-dependent supercurrent in graphene has recently been reported by other groups [18, 19].

The observation of gate tunable supercurrent sets the SGS junctions apart from conventional JJs, whose *V-I* characteristics are typically fixed at a given temperature and magnetic field. In particular, at temperature $T$=0, $I_c$ ~$2\Delta/eR_n$ for diffusive SNS or S/Insulator/S (SIS) junctions, and $I_c$ ~$2\pi e\Delta/h$ for two superconductors coupled via a single quantum channel with perfect transmission[27]. Thus, both $I_c$ and the product $I_cR_n$ are expected to be constants for a given device, and the latter only depend on the energy gap $\Delta$ of superconducting electrodes. However, for SGS junctions, both quantities are gate tunable. In Fig. 3a-b, we plot the observed critical current $I_c$ and product $I_cR_n$ vs $V_g$ for the device shown in Fig. 2c-d. Here we use independently measured values of $R_n(V_g)$ (taken at a small magnetic field) to calculate $I_cR_n$.

Josephson effects in graphene were first calculated for the ballistic case in ref. [21]. At $T=0$, for a wide and short strip of graphene coupled to superconducting electrodes, the values of $I_c$ and $I_cR_n$ are predicted to be

$$I_c = 1.33\frac{e\Delta W}{\hbar\pi L} \quad \text{and} \quad I_cR_n = 2.08\frac{\Delta}{e} \quad (1a)$$

at the Dirac point, and reach asymptotic values of

$$I_c = 1.22\frac{e\Delta}{\hbar\pi}k_F W \quad \text{and} \quad I_cR_n = 2.44\frac{\Delta}{e} \quad (1b)$$

when highly doped. In these expressions, $W$ and $L$ are the channel width and length, respectively, $h$ is the Planck's constant and $k_F$ is the Fermi wavelength.

Our experimental results shown in Fig. 3 qualitatively agree with Eq. (1), *e.g.* $I_c$ and $I_cR_n$ attain their minimum values at the Dirac point, and increase with charge density. However, the agreement fails at the quantitative level. Taking $W/L=10$ and $\Delta=110$ μV extracted from MAR measurements, Eq. (1a) yields $I_c \sim 110$nA at the Dirac point, where the experimental value is observed to be 6 nA. Another important discrepancy is that the $I_cR_n$ product is predicted to have relatively weak dependence on gate voltage – it increases by ~20% from half filling to highly doped regimes, in contrast to the experimentally observed increase of 200-300%. Similar large variations in $I_cR_n$ vs $V_g$ were also reported by other groups [18, 19].

Here we explore the roles played by thermal fluctuations in SGS Josephson junctions. The strong depression of $I_c$ is reminiscent of the behavior of premature switching in underdamped junction described in RCSJ model, in which the junction is shunted by a resistance $R_j$ and a capacitance $C_j$ [23]. For SNS junctions, we can take $R_j \sim R_n$. Within this model, the superconducting phase across the junction has the mechanical analogue of a particle in a tilted washboard potential (Fig. 3c inset), with a frictional force (dissipation) that scales with $1/R_n$. Hence, the bias current corresponds to the tilt of washboard, the superconducting state of the junction corresponds to the particle localized within one of the potential minima, and the resistive state to that rolling continuously down the potential. The RCSJ model is usually parameterized by the "quality factor" $Q$, which is defined by:

$$Q = \omega_p R_j C_j \quad (2)$$

where $\omega_p=(2eI_{c0}/\hbar C)^{1/2}$ is the plasma frequency of the junction and $I_{c0}$ is the intrinsic critical current in the absence of fluctuation. The parameter $Q$ indicates the dissipation experienced by the particle: the junction is overdamped if $Q<<1$, and underdamped if $Q>1$. Experimentally, an underdamped junction can be uniquely identified by its hysteretic *V-I* characteristics, that is, the retrapping current $I_r$ (at which the junction switches from the resistive state to the superconducting state) occurs at a lower value than $I_c$, reflecting the effect of the inertia of a particle moving in a low friction potential. On the other hand, a junction with non-hysteretic *V-I* characteristics may either be overdamped, or under-damped but with weak Josephson coupling or strong thermal fluctuations. Examining our SGS devices, hysteretic *V-I* curves were observed in the majority of the junctions, indicating that most, if not all, of the devices are underdamped. As shown in Fig 3c, we also observe that the hysteretic feature in

SGS junctions is gate tunable: the hysteresis becomes much smaller or even vanishes when $V_g$ tunes graphene from highly doped to zero doping regimes.

In the presence of thermal and other fluctuations, premature switching is expected to occur in underdamped junctions, as the particle is thermally activated over the energy barrier; due to the low damping, once an event of "escape" happens, the particle accelerates down the washboard and is never re-trapped. This premature switching is stochastic and induces a distribution in measured values of $I_c$, the average of which can be much smaller than $I_{c0}$. The mean value of $I_c$ can be approximated by the formula [23]:

$$I_c = I_{c0}\{1 - [(k_B T/2E_J)\ln(\omega_p \Delta t/2\pi)]^{2/3}\} \qquad (3)$$

where $k_B T$ is the thermal energy, $E_J = \hbar I_c/2e$ is the Josephson couple energy and $\Delta t \sim 0.1$s is the sweeping time of the biased current through the dense part of the distribution of observed $I_c$ values. Thus $I_c$ will be significantly reduced from its "intrinsic" value at finite temperature, if thermal fluctuation is non-negligible compared with the Josephson coupling energy. For our devices, $\omega_p \sim 10^{11}$ - $10^{12}$ Hz, so the logarithmic term yields ~ 21-23. At the Dirac point, if we estimate $I_c$=110 nA from Eq. (1a), the ratio $k_B T/2E_J$ is ~0.05. This implies that the critical current is almost completely suppressed, and that the junction's V-I curves is non-hysteretic due to thermal fluctuations, in agreement with experimental observation. When graphene is highly doped, the ratio $k_B T/2E_J$ decreases proportionally with increasing $I_{c0}$, indicating the decreasing importance of thermal energy. Thus premature switching can readily account for the much-reduced values of $I_c$ observed in all experiments to date, as well as V-I characteristics that are hysteretic at high charge carrier density, and non-hysteretic at the Dirac point.

We now seek to understand the unexpectedly strong dependence of $I_c R_n$ on $V_g$ within the model of premature switching in underdamped Josephson junctions. The gate-dependent $R_n$, which sets SGS apart from other JJs, plays a vital role here. If we assume $I_{c0}R_n \sim$ constant, as implied by Eq. (1), changing gate voltage effectively tunes $I_{c0}$ and all other parameters that depend on $I_{c0}$, including $E_J$, $\omega_p$, and $Q$. Thus, as $V_g$ increases from the Dirac point, $R_n$ decreases, leading to a smaller $k_B T/2E_J$ ratio and a less-suppressed $I_c$. Such relative increase in $I_c$ is more than compensates of the decrease in $R_n$, resulting in a larger $I_c R_n$ product.

Quantitatively, we assume $I_{c0}R_n$=constant from Eq. (1), and calculate $I_{c0}$ using measured values of $R_n(V_g)$; the values of $I_c(V_g)$ in the presence of thermal fluctuation are then computed using Eq. (3), and multiplied by $R_n$. The resulting $I_c R_n(V_g)$ are normalized to the value at the Dirac point (hence the exact value of the constant is inconsequential), and shown in Fig. 4a. For comparison, normalized data (Fig. 4b) shown in Fig 4b. In the simulation, between the Dirac point and the highly doped regimes, $I_c R_n$ varies by a factor of ~3-3.5, in reasonable agreement with the factor of 2.2-2.8 observed in the data. We note that our simple simulation assumes $I_c R_n$=constant, and does not take into account the 20% variation predicted by Eq. 1. This can partially explain the larger variation in $I_c R$ in the simulation. Moreover, although our calculation is motivated by theoretical predictions for transport in the ballistic regime, both the RCSJ model and the assumption of a constant, "intrinsic" $I_cR_n$ product are quite general; hence our results should have wide applicability. Finally, the gate dependence of the junction's resistance leads to similar tunable behavior in its quality factor $Q$ or dissipation. This is similar to the gate-tunable dissipation observed in carbon nanotube JJ [28]. Because of its relatively large range of tunable $R_n$, graphene can be used to provide a tunable shunt resistor in other JJ for study of dissipation and quantum coherence [29].

In conclusion, our observation of depression of critical current $I_c$ and the strong dependence of $I_cR_n$ on charge density of SGS junctions can be satisfactorily accounted for by premature switching in underdamped Josephson junctions.

**Acknowledgement**

We thank Gil Refael, Marc Bockrath and Victor Galitski for stimulating discussions and Gang Liu for assistance with experiments. This research is supported in part by NSF CAREER DMR/0748910 and ONR/DMEA Award H94003-07-2-0703.

Fig.1. (a) SEM image of a graphene Josephson transistor device; (b) Differential conductance vs. voltage showing MAR peaks

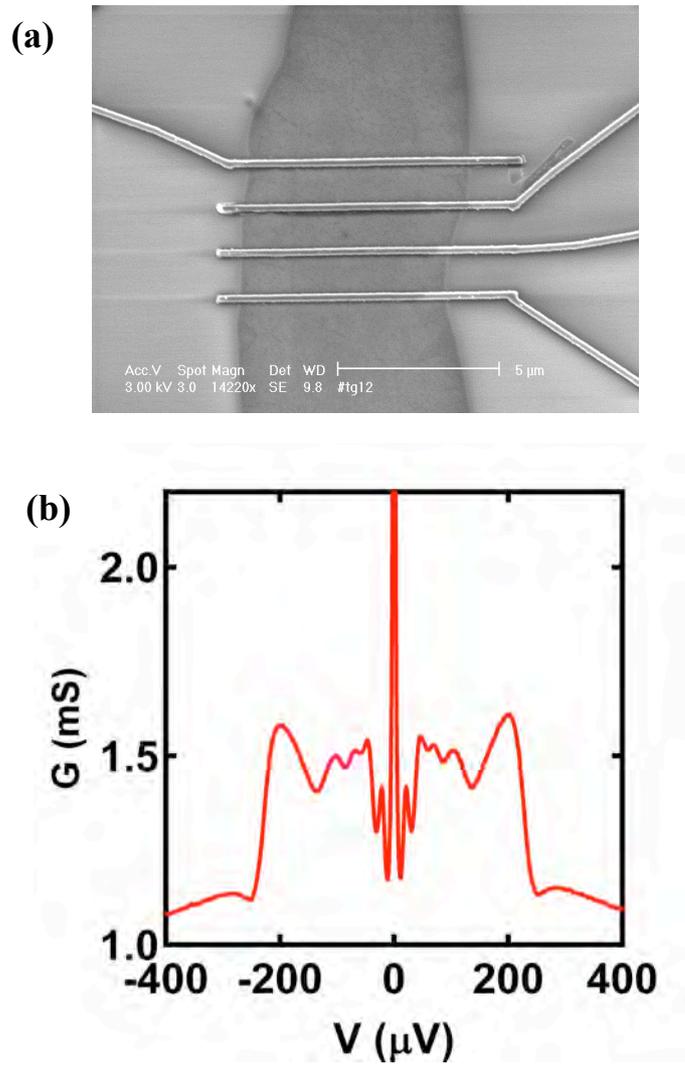

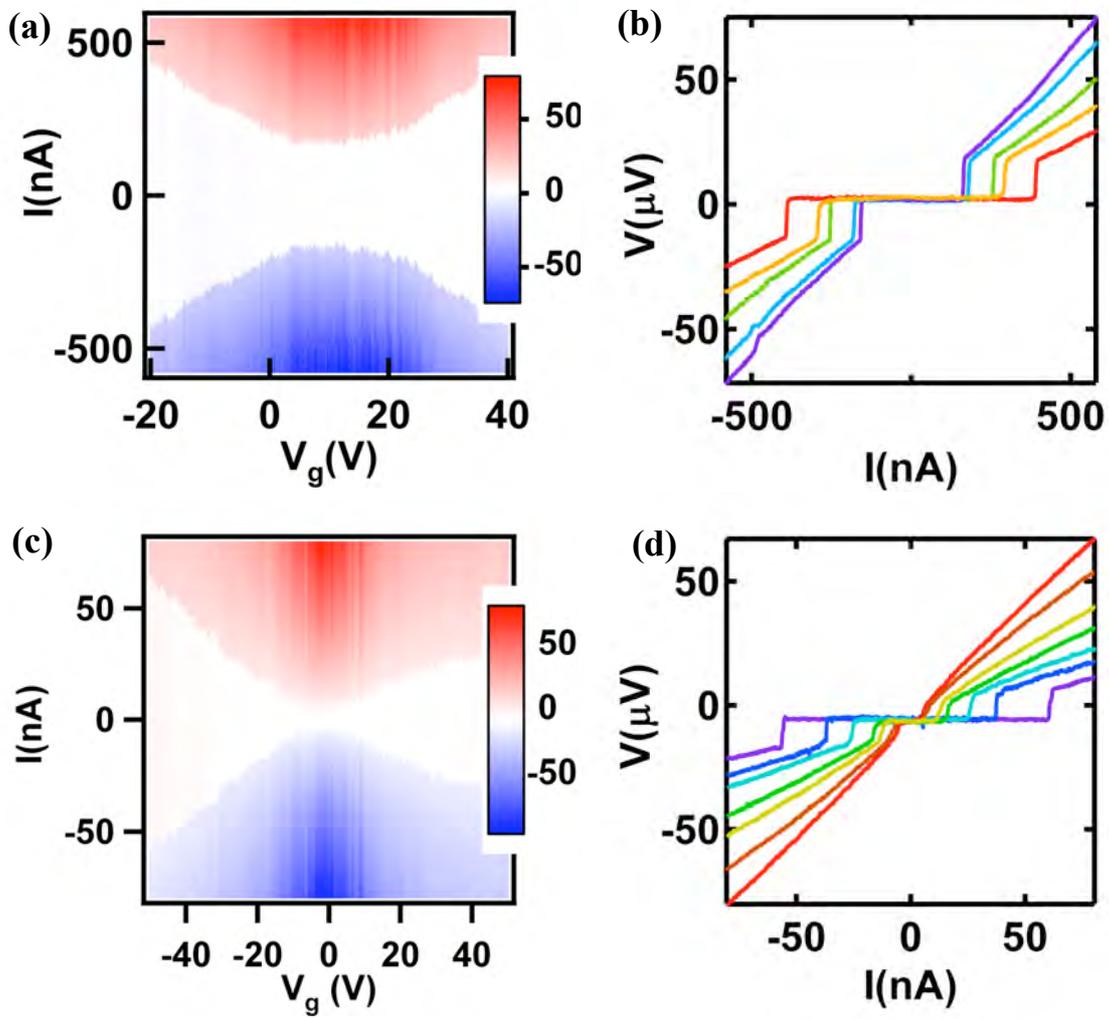

Fig.2. (a) Voltage across the junction as a function of biased current (y-axis) and gate voltage (x-axis). The color scale of voltage is in units of µV; (b) *V-I* line traces at different gate voltages taken from Fig 1(a); (c) (d) Similar data sets from another device.

Fig.3. (a) Critical current $I_c$ vs. gate voltage showing bi-polar supercurrent; (b) Product of $I_cR_N$ as a function of gate voltage for the same device; (c) V-I curves at different gate voltages showing gate tunable hysteresis observed in SGS junctions. Inset: sketch of a "tilted washboard" potential and localized particle described in the RCSJ model.

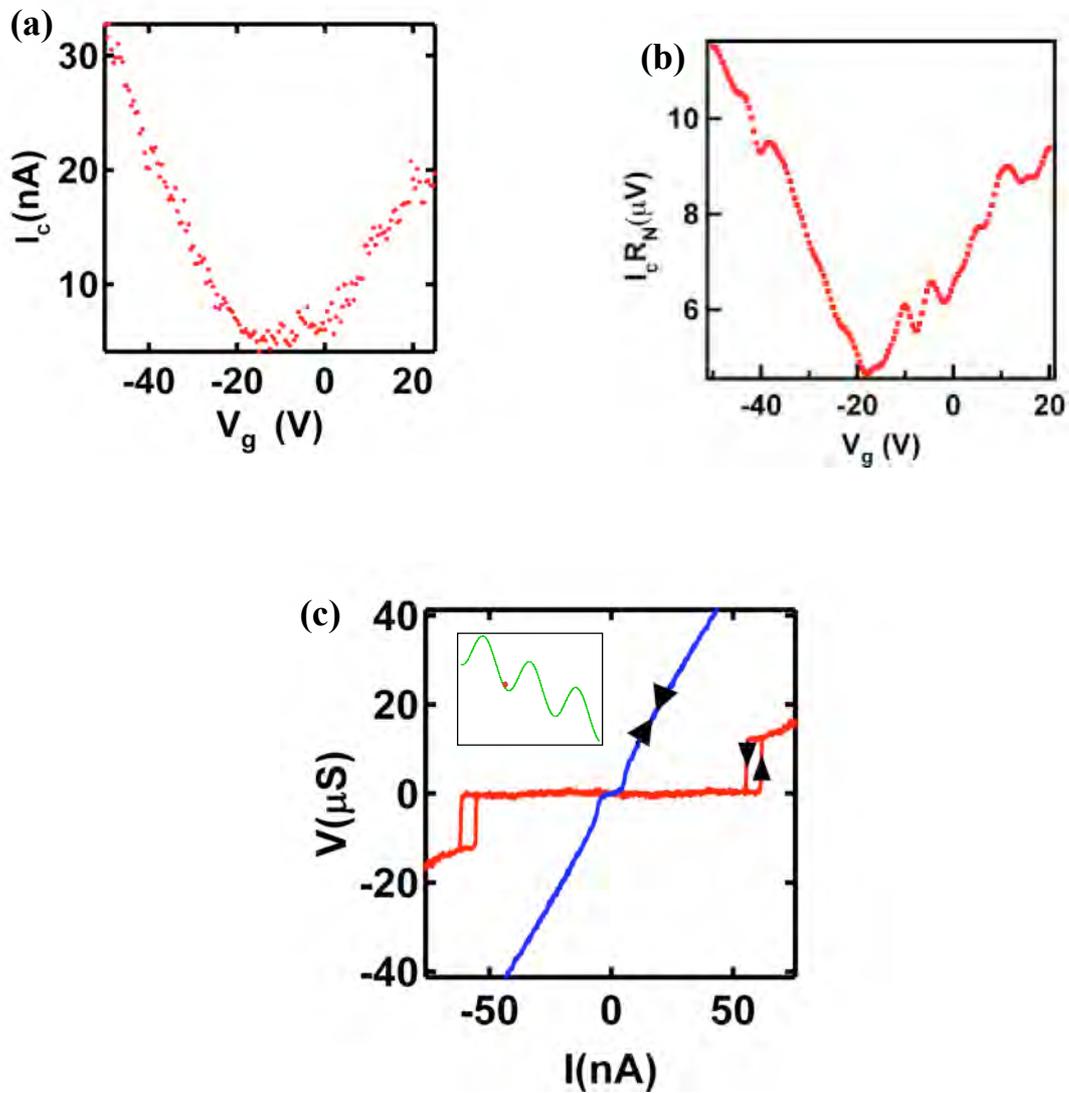

Fig.4. (a) Calculated and (b) experimental curves of normalized $I_cR_N$ vs. $V_g$, where $I_cR_N$ is normalized to the value at the Dirac point.

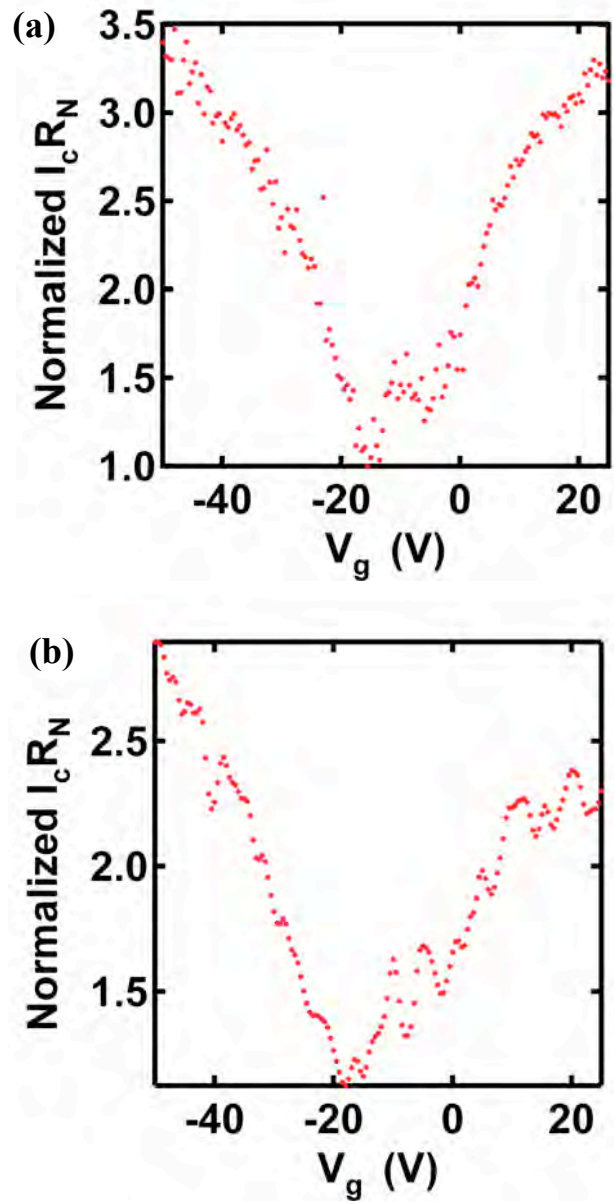